\documentclass[11pt]{article}
\usepackage{epsfig}
\usepackage{dcolumn}% Align table columns on decimal point
\usepackage{bm}% bold math
\usepackage{graphicx}% Include figure files
\usepackage[textsize=tiny]{todonotes}
\usepackage{subcaption}
\usepackage{here}
\usepackage{amssymb,amsmath}
\usepackage{multirow}
\usepackage{cite,color,url}
\usepackage{tabu}
\usepackage{array}
\usepackage[colorlinks=true,urlcolor=teal,anchorcolor=blue
,citecolor=teal,filecolor=blue,linkcolor=teal,menucolor=blue
,linktocpage=true,pdfproducer=medialab,pdfa=true]{hyperref}

\usepackage{colordvi}
\usepackage{epsfig,psfrag,rotating,soul}
\def\beqa{\begin{eqnarray}}
\def\eeqa{\end{eqnarray}}

\newcommand{\htb}[1]{{\color{black} #1}}	

\evensidemargin \oddsidemargin
\allowdisplaybreaks
\renewcommand{\arraystretch}{1.5}
\psfrag{lk}{$l_k$}
\psfrag{lm}{$l_m$}
\psfrag{blk}{$\bar {l}_k$}
\psfrag{blm}{$\bar {l}_m$}
\def\thefootnote{\fnsymbol{footnote}}
\let\OLDthebibliography\thebibliography
\renewcommand\thebibliography[1]{
	\OLDthebibliography{#1}
	\setlength{\parskip}{0pt}
	\setlength{\itemsep}{0pt plus 0.3ex}}
\usepackage{makecell}

\begin{document}

	\thispagestyle{empty}
	\begin{center}
		\begin{Large} %\color{teal}
			\textbf{\textsc{Accommodating the LHC Charged Higgs Boson Excess \\\vspace{0.2cm} at 130 GeV in the General Two-Higgs Doublet Model}}
		\end{Large}
		
		\vspace{1cm}
		
		Abdesslam Arhrib,$^{1,2}$% 
		\footnote{\tt
			%\href{aarhrib@gmail.com}
			{aarhrib@gmail.com}}
		Mohamed Krab$^{3}$% 
		\footnote{\tt
			%\href{mkrab@hep1.phys.ntu.edu.tw}
			{mkrab@hep1.phys.ntu.edu.tw}}
		and Souad Semlali$^{4,5}$% 
		\footnote{\tt
			%\href{S.Semlali@soton.ac.ma}
			{S.Semlali@soton.ac.uk}}
		
		\vspace*{.7cm}
		
		{\sl
			$^1$Abdelmalek Essaadi University, Faculty of Sciences and Techniques,
			B.P. 2117 Tétouan, Tanger, Morocco}\\
		\vspace*{0.1cm}
		{\sl
			$^2$Department of Physics and Center for Theory and Computation, National Tsing Hua University, Hsinchu 300, Taiwan}\\
		\vspace*{0.1cm}
		{\sl
			$^3$Department of Physics, National Taiwan University, Taipei 10617, Taiwan}\\
		
		\vspace*{0.1cm}
		{\sl
			$^4$School of Physics and Astronomy, University of Southampton,
			Southampton SO17 1BJ, UK\\
			\vspace*{0.1cm}
			$^5$Particle Physics Department, Rutherford Appleton Laboratory, Chilton, Didcot, Oxon OX11 0QX, UK}
	\end{center}
	
	\vspace*{0.1cm}

	\begin{abstract}
		Charged Higgs bosons are common predictions in most extensions of the Standard Model (SM) Higgs sector. Therefore, their observation would elucidate the nature of the Higgs sector. 
		Motivated by the ATLAS collaboration's latest analysis performed with $139~\text{fb}^{-1}$ of Run 2 data intended to search for charged Higgs boson, produced in top quark decay and subsequently decaying via $H^\pm \rightarrow cb$, where an excess with a local significance of $3\sigma$ is observed at $m_{H^\pm} = 130~\rm{GeV}$, we discuss here the possibility of explaining such excess in the context of the general 2-Higgs Doublet Model (2HDM type-III), after satisfying all theoretical and up-to-date experimental constraints. We also propose phenomenological scenarios to further explore the mass region around 130 GeV in the four Yukawa types of the 2HDM type-III and suggest alternative decay channel $H^\pm \rightarrow cs$ and/or $H^\pm \rightarrow W^{\pm *}h$ to probe the nature of the observed excess (if it is not a statistical fluctuation). Future searches for $H^\pm$ will be critical in confirming or refuting the first hint of a light charged Higgs boson at the LHC.  
		
	\end{abstract}
	
	\def\thefootnote{\arabic{footnote}}
	\setcounter{page}{0}
	\setcounter{footnote}{0}
	\newpage
	\section{Introduction}
	%%%%%%%
	%%%%%%%%
	\label{intro}
	
	The Large Hadron Collider (LHC) holds the promise of unravelling the mysteries surrounding electroweak symmetry breaking (EWSB) in the near future with its broad Higgs physics program and its high luminosity (HL). Over the past decade, an extensive campaign has been initiated following the Higgs discovery\cite{ATLAS:2012yve,CMS:2012qbp} to measure its properties~\cite{ATLAS:2022vkf,CMS:2022dwd}. This concerted effort has not only spawned new ideas throughout the program's completion but also prompted an alternation of our attention from observations to precision measurements of various properties. The large dataset of the HL-LHC is anticipated to bring about a transformative phase in the realm of Higgs physics, enabling a remarkable accuracy of the Higgs properties and reaching percent-level precision across most of the channels~\cite{Dainese:2019rgk,ATLAS:2022hsp}.
	
	Meanwhile, the possibility of new physics beyond the minimal Higgs framework of the Standard Model (SM) is greatly motivated. Among the various extensions explored, one of the simplest and extensively investigated is the 2-Higgs Doublet Model (2HDM), where the Higgs sector involves new scalars, e.g. an extra CP-even besides the one similar to the SM, one CP-odd and a pair of charged Higgs. In the Yukawa sector, the presence of an additional doublet introduces an interesting phenomenology due to the distinct and large number of Higgs-fermion interactions. A further consequence is the presence of tree level Flavour Changing Neutral Currents (FCNCs) mediated by neutral and/or charged scalars, which, however, are highly suppressed by various experiments. There are in fact different mechanisms aimed at eliminating the unwanted FCNCs or at least restricting their presence, for example implementing the alignment of the Yukawa couplings in flavour space (abbreviated as \textquoteleft A2HDM\textquoteright)~\cite{Pich:2009sp}, or introducing a global discrete $Z_2$ symmetry to allow only one Higgs doublet to couple to all fermions with the same electric charge, and thus forbidding the presence of non-diagonal Yukawa couplings. The different transformations of the quarks' ﬁelds under $Z_2$ symmetry lead to four different 2HDM realisations~\cite{Aoki:2009ha}, named type-I (one of the doublets couples to all fermions), type-II (one doublet couples to up quarks while the second one couples to down quarks), type-X (or Lepton speciﬁc where one of the doublet couples to all quarks while the other couples to leptons) and type-Y (or Flipped where one of the doublet couples to up quarks and leptons whereas the second doublet couples to down quarks). Another suggestion is to implement a flavour symmetry, which guarantees a certain form of the Yukawa matrices (using a Hermitian four-zero
	texture for the Yukawa matrices~\cite{Fritzsch:2002ga,Diaz-Cruz:2004hrt}), and requires the non-diagonal Yukawa terms to obey the following pattern $g_{ij} \propto \sqrt{m_im_j}\chi_{ij}^f$ ({{\it Cheng-Sher ansatz}})~\htb{\cite{Cheng:1987rs,Atwood:1996vj,Sher:2022aaa}}. This generic form of 2HDM with tree level FCNCs (referred here as \textquoteleft2HDM type-III\textquoteright) is the focus of the present work.   
	
	2HDM type-III has garnered a lot of attention for its ability to resolve the experimental tensions in semi-leptonic B decays, e.g. $B\to D^{(*)}\tau \nu,~B \to K^{(*)}\mu\mu$, and $B_\mu \to \tau \nu$\cite{Crivellin:2013wna,Akeroyd:2017mhr,Arhrib:2017yby,Chen:2018hqy}, explain the observed kaon direction CP-violation $Re(\epsilon_K\prime/\epsilon_K)$\cite{Chen:2018ytc} and address the anomalous muon magnetic moment $(g-2)_\mu$  besides other lepton-flavour violating processes~\cite{Omura:2015nja,Omura:2015xcg,Benbrik:2015evd,Iguro:2018qzf,Hou:2021sfl}.  This model is also highlighted for successfully accommodating the excesses~\cite{Benbrik:2022azi,Belyaev:2023xnv} reported in the light Higgs-boson searches within the sub-100 GeV range in the di-photon~\cite{CMS:2023yay,ATLAS:2018xad}, di-tau~\cite{CMS:2022goy} and $bb$\cite{ALEPH:2006tnd} channels. Moreover, the involvement of new parameters in charged Higgs couplings leads to relaxing the constraint from $b\to s\gamma$ on its mass. Unlike in the type-II framework, where a lower bound of $580$ GeV is placed on $m_{H^\pm}$ at 95\%C.L~\cite{Misiak:2017bgg}, 2HDM type-III opens up the possibility for a lighter charged Higgs. 
	
	The presence of a charged Higgs is indeed a striking feature within the 2HDM since different low- and high-energy observables display a strong sensitivity to its contribution. Here, we will turn to the phenomenological implications of the Yukawa texture in searching for light charged Higgs boson at high-energy colliders. 
	
	Previously, the LEP collaborations performed a combination of the searches for pair-produced charged Higgs bosons in Drell-Yan events,  e.g. $e^+e^- \to \gamma/Z \to H^+H^-$, in the $\tau\nu$ and $cs$ final states~\cite{ALEPH:2013htx}. The combined results set a lower limit of 80 GeV on $m_{H^\pm}$ at 95\% C.L under the assumption that the fermionic decays dominate the charged Higgs decay width. In the framework of 2HDM  type-I, the limit on charged Higgs mass is slightly weakened, $m_{H^\pm} > 72~\text{GeV}$, if the bosonic decay channels are open, e.g. $H^\pm  \to W^\pm \phi$, with $m_\phi>12~\text{GeV}$~\cite{ALEPH:2013htx}. 
	
	At hadron colliders, both ATLAS and CMS Collaborations searched for low-mass charged Higgs bosons ($m_{H^\pm}<m_{\text{top}}$) at Run 1 in the final states $\tau \nu$~\cite{ATLAS:2012tny,ATLAS:2012nhc,CMS:2012fgz,ATLAS:2014otc,CMS:2015lsf}, $cs$~\cite{ATLAS:2013uxj,CMS:2015yvc} and $cb$\cite{CMS:2018dzl}. The main $H^\pm$ production mode at the LHC is through the top quark decay, in a double-resonant top production, e.g. $pp \rightarrow t\bar t \rightarrow b\bar{b} H^\pm W^\mp$. At 13 TeV, additional searches for $H^+ \to \tau^+ \nu\text{\cite{ATLAS:2018gfm,CMS:2019bfg}},~cs\text{\cite{CMS:2020osd}},~W^+ A ~\text{\cite{CMS:2019idx}}$  were carried out using data collected with an integrated luminosity of $35.9~\text{fb}^{-1}$. These experiments have placed upper limits on charged Higgs decay rates, e.g. $\mathrm{BR}(t\to H^+b)\times \mathrm{BR}(H^+ \to \tau^+ \nu)\leqslant 1.5\times 10^{-3}$~\cite{ATLAS:2018gfm,CMS:2019bfg} and $\mathrm{BR}(t\to H^+b)\times \mathrm{BR}(H^+ \to c\bar b+c\bar s) \leqslant2.7\times 10^{-3}$~\cite{CMS:2020osd}. More recently, the ATLAS collaboration reported their search for $H^+ \to c\bar b$~\cite{ATLAS:2023bzb} in $t\bar{t}$ events, which benefits from the reduced contribution of the irreducible SM background, $t \to W^+(\to c\bar b)b$, due to the suppression arising from the small Cabibbo-Kobayashi-Maskawa matrix element ($|V_{cb}|)$), with $139~\text{fb}^{-1}$ data, where an excess with a local (global) significance of $3\sigma$ $(1.6\sigma)$  is observed at $m_{H^\pm} = 130~\rm{GeV}$, with a best-fit on $\mathrm{BR}(t \rightarrow H^+ b)\times \mathrm{BR}(H^+ \rightarrow c\bar b) = (0.16 \pm 0.06)\%$ \cite{ATLAS:2023bzb}. Such large branching ratio $\mathrm{BR}(H^+ \to cb)$ was  predicted by several models, such as 2HDM type-III~\cite{Hernandez-Sanchez:2012vxa}, \htb{multi}-Higgs Doublet models, e.g. Three-Higgs Doublet Model (3HDM) without Natural Flavour Conservation (NFC)~\cite{Akeroyd:2016ssd,Ivanov:2021pnr}~\textendash~\htb{3HDM with $Z_2$ symmetry~\cite{Akeroyd:2016ssd}, 3HDM with higher-order CP-symmetry in the Yukawa and scalar sector~\cite{Ivanov:2021pnr}}~\textendash~and 3HDMs with NFC~\cite{Akeroyd:2018axd,Akeroyd:2022ouy}. Recently, the possibility of explaining the 130 GeV excess within the 3HDM and the A2HDM has been discussed in~\cite{Akeroyd:2022ouy,Bernal:2023aai}. This is in stark contrast to the 2HDM with NFC, where none of its four versions (type-I, -II, -X, and -Y) can account for the observed 130 excess, as discussed in ~\cite{Akeroyd:2022ouy,Bernal:2023aai}. The inability of type-I and type-X is attributed to the small branching ratio of $H^+ \to c\bar b$, while in type-II and type-Y, the charged Higgs mass is restricted to be above 580 GeV due to the $b \to s\gamma$ constraint.
	
	%Motivated by the previous results outlined in~\cite{Hernandez-Sanchez:2012vxa}, 
	In this study, we aim to explore the 2HDM type-III parameter space while taking advantage of various re-definitions of the model along with the distinctive charged Higgs couplings, to explain the 130 GeV excess and investigate the implication of such excess on the charged Higgs phenomenology at the LHC. 
	
	This paper is organised as follows. In section~\ref{2hdm_model} we give an overview of the 2HDM with particular Yukawa textures. In section~\ref{sect:constr} we discuss a whole set of theoretical and experimental constraints that should be satisfied by each point in our parameter space. In section~\ref{results} we highlight our results, and then we conclude.

	\section{2HDM type-III with a \textquoteleft specific\textquoteright \hspace*{0.07cm} Yukawa texture}
	\label{2hdm_model}
	The 2HDM is one of the simplest extensions of the SM with an extra $SU(2)_L$ doublet of hypercharge $Y=+1$. The most general, renormalisable, scalar potential is typically written for two Higgs doublets, $\Phi_{1,2}=(\phi^\pm_{1,2},\phi^0_{1,2})^T$, in a generic basis as follows:
	\begin{eqnarray}
	V_{\rm{Higgs}}(\Phi_1,\Phi_2) &=& \lambda_1(\Phi_1^\dagger\Phi_1)^2 +
	\lambda_2(\Phi_2^\dagger\Phi_2)^2 +
	\lambda_3(\Phi_1^\dagger\Phi_1)(\Phi_2^\dagger\Phi_2) +
	\lambda_4(\Phi_1^\dagger\Phi_2)(\Phi_2^\dagger\Phi_1) \nonumber \\ && +
	\left[\frac{1}{2} \lambda_5(\Phi_1^\dagger\Phi_2)^2 + (\lambda_6 (\Phi_1^\dagger \Phi_1) + \lambda_7 (\Phi_2^\dagger\Phi_2))(\Phi_1^\dagger\Phi_2)+\rm{h.c.}\right] \nonumber \\ && 
	+\,m_{11}^2 \Phi_1^\dagger \Phi_1+ m_{22}^2\Phi_2^\dagger
	\Phi_2 - \left[m_{12}^2
	\Phi_1^\dagger \Phi_2 + \rm{h.c.}\right].  \label{2hdmpot}
	\end{eqnarray}
	Restricting ourselves to the case with no CP-violation in the 2HDM scalar sector, all (squared) mass parameters, namely $m_{11}^2$, $m_{22}^2$, and $m_{12}^2$ along with the other dimensionless quartic couplings $\lambda_{1-7}$, are assumed to be real-valued. In this general scheme of the 2HDM with a particular Yukawa texture, a flavour symmetry is implemented to restrict the tree level Higgs mediated FCNCs instead of employing an extra global $Z_2$ symmetry which removes both $\lambda_6$ and $\lambda_7$.\footnote{In the 2HDM with a four-zero Yukawa texture, the presence of $\lambda_6$ and $\lambda_7$ is necessary to ensure the decoupling of the other heavy scalars ($m_H,~m_A,~m_{H^+}\gg v$ while $m_h\sim \mathcal{O}(v)$)~\cite{Gunion:2002zf}. This is not relevant here where all the physical masses are lying below the Electro-Weak (EW) scale, we, therefore, set $\lambda_6=\lambda_7 =0$ from the Higgs potential given by Eq.~(\ref{2hdmpot}).}
	
	After the EWSB, the scalar sector involves five physical Higgses: two CP-even, $h$ and $H$ with $m_h<m_H$, one CP-odd ($A$) and a pair of charged Higgs ($H^\pm$). One of the CP-even can be identified as the observed Higgs at the LHC with properties that approach those of the SM-like. \\
	One can describe the 2HDM \htb{in the physical basis}, using the following set of parameters:
	\begin{eqnarray}
	m_h,~~m_H,~~m_A,~~m_{H^\pm},~~\sin(\beta-\alpha),~~\tan \beta=v_2/v_1~~\text{and}~~m_{12}^2,
	\end{eqnarray} 
	where $\alpha$ is the mixing angle in the CP-even sector and $v_{1,2}$ are the vacuum expectation values of the two doublets, with $v_1/v=c_\beta$, $v_2/v=s_\beta$ and $v=\sqrt{v_1^2+v_2^2}$. $c_x$ \htb{denotes $\cos x$ and $s_x$ denotes $\sin x$.}
	
	The couplings of the Higgs bosons to both leptons and quarks are modified due to the mixing in the Higgs sector and the new contribution emerging from the on-diagonal terms of a four-zero Yukawa texture. This results in a distinct and interesting phenomenology w.r.t the SM. We will now move to exploring the Yukawa interactions in the generic 2HDM with tree-level FCNCs (2HDM type-III). The Yukawa Lagrangian where both Higgs doublets couple to all fermions is written as:
		\begin{eqnarray}
		-{\cal{L}}_{Y} & = & 
		\bar{Q}_L^0 Y^{u}_1 {\tilde \Phi_{1}} u^0_{R} +
		\bar{Q}_L^0 Y^{u}_2 {\tilde \Phi_{2}} u^0_{R} +
		\bar{Q}_L^0 Y^{d}_1 \Phi_{1} d^0_{R}  \nonumber \\
		&& +\,\bar{Q}_L^0  Y^{d}_2  \Phi_{2}d^0_{R}  
		+ \bar{L}^0_{L} Y^{{l}}_{1}\Phi_{1}l^0_{R}
		+ \bar{L}^0_{L} Y^{{l}}_{2}\Phi_{2}l^0_{R}, 
		\label{lag-f}
		\end{eqnarray}
		where $Q_L^0$ $(L^0_L)$ are left-handed quark (lepton) doublets, $f^0_R= (u^0_R,~d^0_R,~l^0_R)$ denotes right-handed fermion singlets, and $Y_{1,~2}^{u,~d}~(Y_{1,~2}^l)$ are the ($3\times 3$) Yukawa matrices for quarks (leptons) in flavour space. The superscript "0" indicates that the interaction basis states ($Q_L^0,~u^0_R,~d^0_R$ ) are vectors in flavor space.
		After EWSB, the fermion mass matrix can be expressed as: 
		\begin{eqnarray}
		M_ f= \frac{v}{\sqrt{2}} (c_\beta Y_1^f + s_\beta Y_2^f ),~~ f=u,~d,~l.
		\end{eqnarray}
		Both Yukawa matrices, $Y_1^f$ and $Y_2^f$, can not be diagonalized simultaneously without assuming a correlation between $Y_1^f$ and $Y_2^f$, thus the Lagrangian would exhibit Higgs mediated FCNCs at tree level. The diagonalization of the fermion mass matrices is performed using the bi-unitary matrices in the following way: $f_{L,R}=V_{L,R}^f f_{L,R}^0$, where $f_L$ indicates the physical mass eigenstates of the fermions, $\bar{M}_{f} = V_L^{f{\dagger}} M_f V_R^f$ and $\tilde{Y}_{1,2}^f = V_L^{f\dagger} Y_{1,2}^f V_R^f$. One can expand the Lagrangian in Eq.~\ref{lag-f}, and express the scalar fields $\Phi_i$ in terms of the physical Higgs states,
		\begin{align}
		\phi^\pm_1 &= c_\beta G^\pm  - s_\beta H^\pm, \\ 
		\phi^\pm_2 &= s_\beta G^\pm  + c_\beta H^\pm, \\ 
		\phi^0_1 &= \frac{1}{\sqrt{2}}(v_1 -s_\alpha h+ c_\alpha H + ic_\beta G^0-i s_\beta A),  \\
		\phi^0_2 &= \frac{1}{\sqrt{2}}(v_2 +c_\alpha h+ s_\alpha H + is_\beta G^0+i c_\beta A).
		\end{align}
		% Previously, the authors of Ref.~\cite{Hernandez-Sanchez:2012vxa} {\htr{expressed the Yukawa Lagrangian in a convenient and compact form}}:
		
		The resulting Yukawa Lagrangian can be expressed in a convenient and compact form in terms of the Higgs-fermion interactions, $g_{\phi \bar{f}_i f_j}$, as follows~\cite{Hernandez-Sanchez:2012vxa}:
		\begin{eqnarray}
		-{\cal L}_{\phi \bar{f}_i f_j } & =& g_{h \bar{f_i}f_j} \bar{f}_i  f_j h^0 + g_{H \bar{f_i}f_j} \bar{f}_i  f_j H - i g_{A \bar{f_i}f_j} \bar{f}_i   \gamma_5 A f_j 	\nonumber \\
		&+ & g_{H^+\bar{u_i}d_j} \overline{u}_i d_j \,H^+
		+ \overline{\nu_{i}}~ g_{H^+\bar{l_i}\nu_{l_j}}    l_j H^+
		+\rm{h.c.},  \label{lagrangian1}
		\end{eqnarray}
		where $i$ and $j$ are flavour indices,  $g_{h \bar{f_i}f_j}$, $g_{H \bar{f_i}f_j}$ and $g_{A \bar{f_i}f_j}$ are the  neutral Higgs-fermion-fermion couplings. $g_{H^+\bar{u_i}d_j}$ and $g_{H^+\bar{l_i}\nu_{l_j}}$ are the charged Higgs couplings  $H^+\bar{u_i}d_j$ and $l_i^-\nu_{l_j} H^+$ and, respectively. 
		
		To compare the new physics emerging from the four-zero Yukawa texture with more conventional Two-Higgs-Doublet Models (2HDMs), specifically the 2HDM type-II, previous studies have adopted the following redefinition for the Yukawa couplings, $g_{\phi \bar{f_i}f_j}$, 
		\begin{eqnarray}
		g_{\phi\bar{f_i}f_j}^{\text{2HDM-III}} = g_{\phi \bar{f_i}f_j}^{\text{2HDM-II}} + \Delta g_{\phi \bar{f_i}f_j},
		\end{eqnarray} 
		In this expression, the first term, $g_{\bar{f_i}f_j\phi}^{\text{2HDM-III}}$, represents the deviation of the 'traditional' $Z_2$ symmetric 2HDM-II from the Standard Model (SM) result, while the second term, $\Delta g_{\bar{f_i}f_j\phi}$, captures the new contributions introduced by the 2HDM-III due to the Yukawa texture. These redefinitions, however, are not unique, one can similarly reproduce other variants of the 2HDM, including 2HDM-I, 2HDM-X, or 2HDM-Y. In each case, the coupling in the 2HDM-III can be expressed as: 
		\begin{eqnarray}
		g_{\phi \bar{f_i}f_j}^{\text{2HDM-III}} = g_{\phi \bar{f_i}f_j}^{\text{2HDM-any type}} + \Delta g_{\phi \bar{f_i}f_j},
		\end{eqnarray} 
		demonstrating the flexibility in adapting the 2HDM-III to different 2HDM scenarios. Here, we refer to type-I, II, X, and Y-like versions where the 2HDM type-III is considered as a deviation from the respective 2HDM type-I, II, X, and Y. One can recover 2HDM when $\Delta g_{\bar{f_i}f_j\phi} \to 0$.\\ Adopting the {\it{Chen-Sher ansatz}} in the fermionic sector, which assumes a hierarchy among the Yukawa matrices, results in a suppression of the off-diagonal elements due to the quark mass hierarchy, especially for the first generation where the experimental constraints are the most stringent. As a result, the Higgs Yukawa couplings obey a particular pattern  with the new physics contribution from the Yukawa texture taking the following form~\cite{Cheng:1987rs,Diaz-Cruz:2004hrt}:
		\begin{eqnarray}
		\Delta g_{\bar{f_i}f_j\phi} \sim \frac{\sqrt{m_i m_j}}{v} \chi_{ij}^f,
		\end{eqnarray}  
		where $\chi_{ij}$ are free dimensionless arbitrary parameters which describe the new source of tree level FCNCs.  As outlined above, these parameters are bounded by constraints from flavour physics and colliders experiments ($\chi_{ij}^f\lesssim \mathcal{O}(1)$). The parameters $\chi_{ij}$ can be complex, thereby allowing a new source of CP violation. In this analysis, we assume that the only source of CP-violation is the Cabbibo-Kobayashi-Maskawa (CKM) matrix and that all Higgs-fermion interactions respect CP-invariance. A recent review~\cite{Babu:2018uik} has proposed a new {{\it ansatz}} with a modified Yukawa coupling, e.g. $\Delta g_{\bar{f_i}f_j\phi} \sim  \chi_{ij}^f \min\{m_i,m_j\}/v$ to overcome the limitation of the {{\it Cheng-Sher ansatz}} for multi-Higgs doublet model stemming from the current experimental data.\\

		The Higgs-fermion-fermion couplings $g_{\phi \bar{f}_i f_j}$ in Eq.~\ref{lagrangian1} are written as follows:
		
		\begin{eqnarray}
		\label{coups1}
		g_{h \bar{f_i}f_j} &=& \frac{m_{f_i}}{v}
		\xi_{h}^{f_{i j}} ,  \quad g_{H \bar{f_i}f_j} = \frac{m_{f_i}}{v}
		\xi_{H}^{f_{i j}},
		\quad  g_{A^0 \bar{f_i}f_j}= \frac{m_{f_i}}{v} \xi_{A}^{f_{i j}}.
		\end{eqnarray}
		Table~\ref{coupIII} summarizes the Yukawa interactions $\xi^{f_{ij}}_{h, H, A}$, in the 2HDM type-III. These couplings are expressed in terms of the parameters $\epsilon_{\phi}^f$, $X$, $Y$, and $Z$.  The specific forms of $\epsilon_{h,H}^f$, $X$, $Y$, and $Z$ in different versions of the 2HDM type-III are provided in Table~\ref{couplings}.
		
		\begin{table}[h!]
			\begin{center}
				\begin{tabular}{c|c|c|c} \hline\hline 
					$\phi$  & $\xi^{u_{ij}}_{\phi}$ &  $\xi^{d_{ij}}_{\phi}$ &  $\xi^{\ell_{ij}}_{\phi}$  \\   \hline
					$h$~ 
					& ~ $  \epsilon_h^u \delta_{ij} - \frac{(\epsilon_H^u+Y \epsilon_h^u)}{\sqrt{2} f(Y)} \sqrt{\frac{m_{u_j}}{m_{u_i}}} \tilde{\chi}_{ij}^u$~
					& ~ $ \epsilon_h^d \delta_{ij} + \frac{(\epsilon_H^d-X \epsilon_h^d)}{\sqrt{2} f(X)} \sqrt{\frac{m_{d_j}}{m_{d_i}}} \tilde{\chi}_{ij}^d$~
					& ~ $\epsilon_h^l \delta_{ij} + \frac{(\epsilon_H^l-Z \epsilon_h^l)}{\sqrt{2} f(Z)} \sqrt{\frac{m_{l_j}}{m_{l_i}}} \tilde{\chi}_{ij}^l$ ~ \\
					$H$~
					& $ \epsilon_H^u \delta_{ij} + \frac{(\epsilon_h^u-Y \epsilon_H^u)}{\sqrt{2} f(Y)} \sqrt{\frac{m_{u_j}}{m_{u_i}}} \tilde{\chi}_{ij}^u $
					& $\epsilon_H^d \delta_{ij} - \frac{(\epsilon_h^d+X \epsilon_H^d)}{\sqrt{2} f(X)} \sqrt{\frac{m_{d_j}}{m_{d_i}}} \tilde{\chi}_{ij}^d $ 
					& $\epsilon_H^l \delta_{ij} - \frac{(\epsilon_h^l+Z \epsilon_H^l)}{\sqrt{2} f(Z)} \sqrt{\frac{m_{l_j}}{m_{l_i}}} \tilde{\chi}_{ij}^d$ \\
					$A$~  
					& $ -Y \delta_{ij} + \frac{f(Y)}{\sqrt{2} } \sqrt{\frac{m_{u_j}}{m_{u_i}}} \tilde{\chi}_{ij}^u $  
					& $ -X \delta_{ij} + \frac{f(X)}{\sqrt{2} } \sqrt{\frac{m_{d_j}}{m_{d_i}}} \tilde{\chi}_{ij}^d$  
					& $-Z \delta_{ij} + \frac{f(Z)}{\sqrt{2} } \sqrt{\frac{m_{l_j}}{m_{l_i}}} \tilde{\chi}_{ij}^d$ \\ \hline \hline 
				\end{tabular}
			\end{center}
			\caption {Yukawa couplings of the $h$, $H$, and $A$ bosons to fermions in the 2HDM type-III~\cite{Hernandez-Sanchez:2012vxa}, with $f(x) = \sqrt{1+x^2}$. $\epsilon_{h,H}^f,~X,~Y$ and $Z$ are given in Tab.~\ref{couplings}.} 
			\label{coupIII}
		\end{table}
		The charged Higgs couplings on the other hand are written as follows,
		\begin{eqnarray}
		g_{H^+\bar{u_i}d_j} = -\frac{\sqrt{2}}{v}
		(m_{d_j}X_{i j} P_R +m_{u_i} Y_{ij}P_L), \quad g_{H^+l_i\nu_{l_j}}= \frac{\sqrt 2m_{{l}_j} }{v} Z_{ij} P_R.
		\end{eqnarray}		
		The matrix elements associated with the charged Higgs couplings with quarks and leptons, $X_{ij},~Y_{ij}$ and $Z_{ij}$~\cite{Hernandez-Sanchez:2012vxa}, are defined as:
		\begin{eqnarray}
		X_{i j} & = &   \sum^3_{l=1}  (V_{\rm CKM})_{il} \bigg[ X \, \frac{m_{d_{l}}}{m_{d_j}} \, \delta_{lj}
		-\frac{f(X)}{\sqrt{2} }  \,\sqrt{\frac{m_{d_l}}{ m_{d_j} }} \, \tilde{\chi}^d_{lj}  \bigg], \label{Xij}
		\\
		Y_{i j} & = &  \sum^3_{l=1}  \bigg[ Y  \, \delta_{il}
		-\frac{f(Y)}{\sqrt{2} }  \,\sqrt{\frac{ m_{u_l}}{m_{u_i}} } \, \tilde{\chi}^u_{il}  \bigg]  (V_{\rm CKM})_{lj}, \label{Yij} \\
		Z_{i j}& = &   \bigg[Z \, \frac{m_{{l}_{i}}}{m_{{l}_j}} \,
		\delta_{ij} -\frac{f(Z)}{\sqrt{2} }  \,\sqrt{\frac{m_{{l}_i}}{m_{{l}_j}}  }
		\, \tilde{\chi}^{l}_{ij}  \bigg].
		\label{Zij}
		\end{eqnarray}
		\begin{table}[h!]
			\centering
			\begin{tabular}{|c|c|c|c|c|c|c|c|c|c|}
				\hline
				2HDM-III& $\epsilon^u_h $  & $\epsilon^d_h $ & $\epsilon^d_{l} $  & $\epsilon^u_H $  & $\epsilon^d_H $ & $\epsilon^{l}_H $ & $X$ &  $Y$ &  $Z$  \\ \hline
				type-I-like~
				& $c_\alpha/s_\beta$ & $c_\alpha/s_\beta$ & $c_\alpha/s_\beta$ 
				& $s_\alpha/s_\beta$ & $s_\alpha/s_\beta$ & $s_\alpha/s_\beta$ &  $-\cot\beta$ & $\cot\beta$ & $-\cot\beta$\\
				type-II-like
				& $c_\alpha/s_\beta$ & $-s_\alpha/c_\beta$ & $-s_\alpha/c_\beta$
				& $s_\alpha/s_\beta$ & $c_\alpha/c_\beta$ & $c_\alpha/c_\beta$ & $\tan\beta$ & $\cot\beta$ & $\tan\beta$\\
				type-X-like
				&  $c_\alpha/s_\beta$ & $c_\alpha/s_\beta$ & $-s_\alpha/c_\beta$
				& $s_\alpha/s_\beta$ & $s_\alpha/s_\beta$ & $c_\alpha/c_\beta$ & $-\cot\beta$ & $\cot\beta$ & $\tan\beta$\\
				type-Y-like
				& $c_\alpha/s_\beta$ & $-s_\alpha/c_\beta$ & $c_\alpha/s_\beta$
				& $s_\alpha/s_\beta$ & $c_\alpha/c_\beta$ & $s_\alpha/s_\beta$ & $\tan\beta$ & $\cot\beta$ & $-\cot\beta$\\
				\hline
			\end{tabular}
			%\end{center}
			\caption{ Parameters $\epsilon^f_\phi$, $X$, $Y$ and $Z$  defined in the Yukawa interactions in the four versions of the 2HDM type-III with a specific Yukawa texture~\cite{Hernandez-Sanchez:2012vxa}.}
			\label{couplings}
		\end{table}
		
		In the context of 2HDM type-III, the charged Higgs couplings depend on more than one parameter, leading to a distinctive phenomenology. This contrasts with the 2HDM with NFC, where these couplings depend solely on $\tan \beta$. As pointed previously, the data from flavour physics experiments, particularly from $B_d^0-\bar{B}_d^0$, enforce stringent constraint on the flavour changing couplings involving the first generation of quarks. Consequently, it is reasonable to assume that the Yukawa couplings involving the $u$- and $d$-quarks are suppressed, $\chi^{uj,~dj}\approx 0$ for $j=1,~2,~3$. {Here, we assume that the off-diagonal elements are zero} ($\chi_{ij}^f =0,~i\neq j$).

	\section{Scan strategy and constraints}
	\label{sect:constr}
	In order to investigate whether the 2HDM type-III can explain the slight observed excess at 130 GeV \cite{ATLAS:2023bzb}, we explore its parameter space using the program \texttt{2HDMC-1.8.0} \cite{Eriksson:2009ws}. Note that the Yukawa couplings have been modified to account for the contribution arising from the on/off-diagonal terms of the Yukawa textures. As stated above, we assume that the off-diagonal terms are zero. We set the mass of $H$ to the most recent measurement\footnote{The most recent measurement of the Higgs boson mass from CMS yields $m_H = 125.08 \pm 0.12 ~\rm{GeV}$ \cite{CMS-PAS-HIG-21-019}.} value with unprecedented precision from the ATLAS LHC Run 1 and 2 combined analysis, $m_H = 125.11~\rm{GeV}$ \cite{ATLAS:2023oaq}, and randomly scan the remaining 2HDM parameters within the ranges shown in Tab. \ref{table:scan}.

	{\renewcommand{\arraystretch}{1.05} %donne la distance entre les lignes%
		{\setlength{\tabcolsep}{0.8cm} %donne la distance entre les collones%
			\begin{table}[h!]
				\centering
				\begin{tabular}{c||c} 
					Parameter &  Scanned Range  \\ \hline 
					$m_h~(\mathrm{GeV})$  & $[65,\,110]$  \\
					$m_H~(\mathrm{GeV})$ & $125.11$  \\
					$m_A~(\mathrm{GeV})$ &  $150$  \\  
					$m_{H^\pm}~(\mathrm{GeV})$ & $[110,\,150]$  \\ 
					$\sin(\beta-\alpha)$ & $[-0.3,\,0.3]$ \\ 
					$\tan\beta$ & $[0.5,\,15]$  \\
					$m_{12}^2~(\mathrm{GeV}^2)$ & $m^{2}_h \sin\beta \cos\beta$  \\
					\quad $\lambda_6=\lambda_7$ & $0$  \\ 
					$\chi^{u,d,l}$ & $[-3,\,3]$ \\\hline  \hline
				\end{tabular}	
				\caption{2HDM type-III input parameters.} \label{table:scan}
			\end{table}
			We then	require each parameter point to fulfil the following constraints: 
			\begin{itemize}
				\item Perturbative unitarity \cite{Kanemura:1993hm, Akeroyd:2000wc, Arhrib:2000is}, perturbativity \cite{Chang:2015goa} and vacuum stability \cite{Barroso:2013awa} utilising the method implemented in \texttt{2HDMC}.
				
				\item Electroweak precision data (EWPD) through the oblique parameters $S$, $T$ and $U$ \cite{Grimus:2007if,Grimus:2008nb} utilising the following best-fit results (for $U=0$) \cite{ParticleDataGroup:2020ssz}:
				\begin{equation}
				S=0.05 \pm 0.08,\quad T=0.09 \pm 0.07.
				\end{equation}
				We calculate $S$ and $T$ in the model using \texttt{2HDMC} and
				require $\chi^2_{ST} \leq 6.18$, \htb{which corresponds to a 95.46\% C.L. in a two-dimensional parameter space in the
					Gaussian limit}, where the correlation factor among these parameters has been carefully taken into account.
				
				\item Flavour physics observables utilising the following experimental results:
				\begin{itemize}
					\item $\mathcal{B}(B \rightarrow X_s \gamma) = (3.32 \pm 0.15) \times 10^{-4}$ \cite{HeavyFlavorAveragingGroup:2022wzx},
					
					\item $\mathcal{B}(B \rightarrow \tau\nu) = (1.06 \pm 0.19) \times 10^{-4}$ \cite{HeavyFlavorAveragingGroup:2022wzx},
					
					\item $\mathcal{B}(B_s \rightarrow \mu^+ \mu^-) = (3.09^{+0.46+0.15}_{-0.43-0.11}) \times 10^{-9}$ \cite{LHCb:2021awg,LHCb:2021vsc},
					
					\item $\mathcal{B}(B_d \rightarrow \mu^+ \mu^-) = (1.20^{+0.83}_{-0.74}\pm 0.14) \times 10^{-10}$ \cite{LHCb:2021awg,LHCb:2021vsc},
					
					\item $\mathcal{B}(D_s \rightarrow \tau\nu) = (5.51 \pm 0.24) \times 10^{-2}$ \cite{HFLAV:2016hnz}.
				\end{itemize} 
				The contribution of the 2HDM to the observables listed above is evaluated using \texttt{SuperIso v4.1} \cite{Mahmoudi:2008tp}, where we only select the parameter points that satisfy the $\chi^2$ restriction at the $2\sigma$ level. Note that we have also checked the limit from $B^0-\bar{B}^0$ mixing based on the analysis of Ref. \cite{Hernandez-Sanchez:2012vxa}.
			\end{itemize}
			Additionally, in order to extract the signal strength and check limits from additional Higgs searches, we make use of the code \texttt{HiggsTools} \cite{Bahl:2022igd}, which incorporates the new versions of the codes \texttt{HiggsBounds} \cite{Bechtle:2020pkv} and \texttt{HiggsSignals} \cite{Bechtle:2020uwn}, we demand all the points passing the aforementioned requirements to fulfil the following experimental constraints:
			
			\begin{itemize}			
				\item Exclusion limits at 95$\%$ confidence level from BSM Higgs searches at collider experiments using \texttt{HiggsBounds}. The latter excludes a parameter point if the theory prediction for the most sensitive channel for one of the BSM scalars is larger than the experimentally observed limit.  
				
				\item Agreement with the measured properties of the SM-like Higgs boson with a mass of 125 GeV making use of \texttt{HiggsSignals}. This code returns $\chi^2$ value that can be used to accept (or reject) a parameter point based on the condition $\Delta\chi^2_{125} = \chi^2 - \chi^2_{\rm{SM}} \leq 6.18$, where $\chi^2_{\rm{SM}} = 151.7$ corresponds to the SM fit result obtained with \texttt{HiggsSignals}.
			\end{itemize}
			
			\htb{Parameter points which are allowed by all constraints listed above are shown in Fig. \ref{fig:allconstr} for all Yukawa textures.
				We first check the parameter points that passed both theoretical restrictions as well as exclusion limits from collider experiments using the old version of $\texttt{HiggsBounds}$ (blue points labelled as ``theoretical''). Constraints from EWPD or $STU$ parameters force 2HDM Higgses to be close in mass. For example, light $h$ with a mass of about $\sim 70$ GeV is not feasible until $m_{H^\pm} \gtrsim 128$ GeV (orange points which satisfied theoretical and EWPD). Flavour physics strongly constrains type-II(Y)-like than type-I(X)-like. It is visible (Fig. \ref{fig:allconstr}) that $h$ masses below $\sim$ 80 GeV are almost excluded in type-II(Y)-like (green points which satisfied theoretical, EWPD and flavour constraints) due to $b \to s\gamma$, which is the most constraining observable that excludes low values of $\tan\beta$, which corresponds to light $h$ with masses below $\sim$ 80 GeV after theoretical constraints. \\
				For type-I-like (type-X-like), such observable excludes low (high) $\tan\beta$ values. Parameter points that passed \texttt{HiggsBounds}/\texttt{HiggsSignals} are shown in red (points which satisfied all constraints). We found that the most sensitive charged Higgs searches allowing parameter points are $H^\pm \to \tau\nu$ \cite{ATLAS:2018gfm} and $H^\pm \to cs+cb$ \cite{CMS:2020osd}. In addition, the parameter space of type-I(X)-like is mostly sensitive to LEP searches for the SM Higgs boson \cite{LEPWorkingGroupforHiggsbosonsearches:2003ing} and LHC search for $pp \to \phi \to \tau\tau$ \cite{CMS:2015mca}, whereas in type-II(Y)-like, the survived points are sensitive to low-mass resonances decaying into a pair of bottom quark \cite{CMS:2018pwl} and $pp \to \phi \to \tau\tau$ \cite{CMS:2015mca}. Note that constraint coming from the total decay width of the top quark ($\Gamma_t = 1.42^{+0.19}_{-0.15}~\rm{GeV}$ \cite{ParticleDataGroup:2020ssz}), which is modified by the presence of the $t \to b H^\pm$ decay, is also checked. Our data are consistent with the measurement at 1$\sigma$ level.}
			
			\begin{figure}[t!]
				\centering
				\includegraphics[scale=0.6]{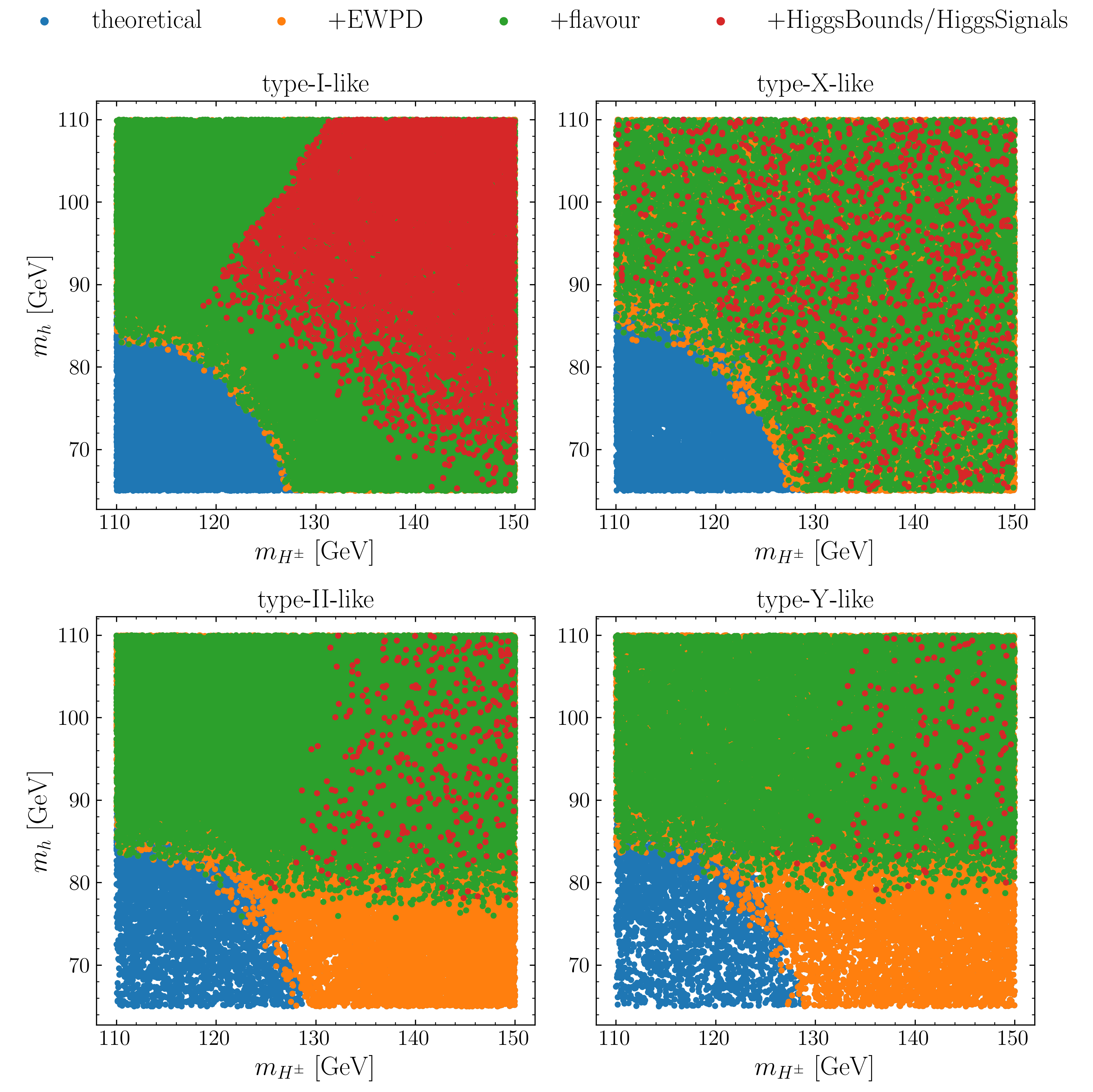}
				\caption{Parameter points which are allowed by theoretical and experimental constraints. Blue points satisfy theoretical restrictions and exclusion limits from collider experiments using the old version of $\texttt{HiggsBounds}$, labelled as ``theoretical''. Orange points satisfy theoretical and EWPD constraints. Green points satisfy theoretical, EWPD, and flavour constraints. Finally, red points satisfy all the constraints.}
				\label{fig:allconstr}
			\end{figure}

			\section{Discussion}
			\label{results}
			As outlined above, the ATLAS group has reported a local excess of 3$\sigma$ at $m_{H^\pm} = 130$ GeV~\cite{ATLAS:2023bzb}. The best-fit $\mathrm{BR}(t \rightarrow H^+ b)\times\mathrm{BR}(H^+ \rightarrow c\bar b)$, corresponding to this measurement, is determined to be ($0.16 \pm 0.06$)\%. In this section, we first discuss the prospects of the four possible Yukawa realisations in light of the aforementioned excess, and the current experimental bounds on charged Higgs decay rates derived from previous searches specifically the limit from $\mathrm{BR}(t\to H^+b)\times \mathrm{BR}(H^+ \to \tau\nu)$ \cite{ATLAS:2018gfm,CMS:2019bfg}.
			%We specifically require that all parameter points are below the limit from $H^\pm \to \tau\nu$ decay~\cite{ATLAS:2018gfm,CMS:2019bfg},
			%\begin{equation}
			%\mathrm{BR}(t\to H^+b)\times \mathrm{BR}(H^+ \to \tau\nu)\leqslant 1.5\times 10^{-3}.
			%\end{equation}  
			We then explore the region around 130 GeV and study the phenomenological consequences at the LHC.
			
			Fig.~\ref{fig1} shows the parameter points in the four Yukawa types of the 2HDM type-III, after imposing constraints of theoretical and experimental nature, for the product of branching ratios $\mathrm{BR}(t \rightarrow H^\pm b)\times\mathrm{BR}(H^\pm \rightarrow cb)$ \htb{in the mass range between 80 and 160 GeV.} The observed (expected) 95\% C.L upper limit on $\mathrm{BR}(t \rightarrow H^\pm b)\times\mathrm{BR}(H^\pm \rightarrow cb)$ vary between 0.15\% (0.009\%) and 0.42\% (0.25\%)  for a charged Higgs in the mass range between 60 and 160 GeV~\cite{ATLAS:2023bzb}.  
			
			\begin{figure}[t!]
				\centering
				\includegraphics[scale=0.6]{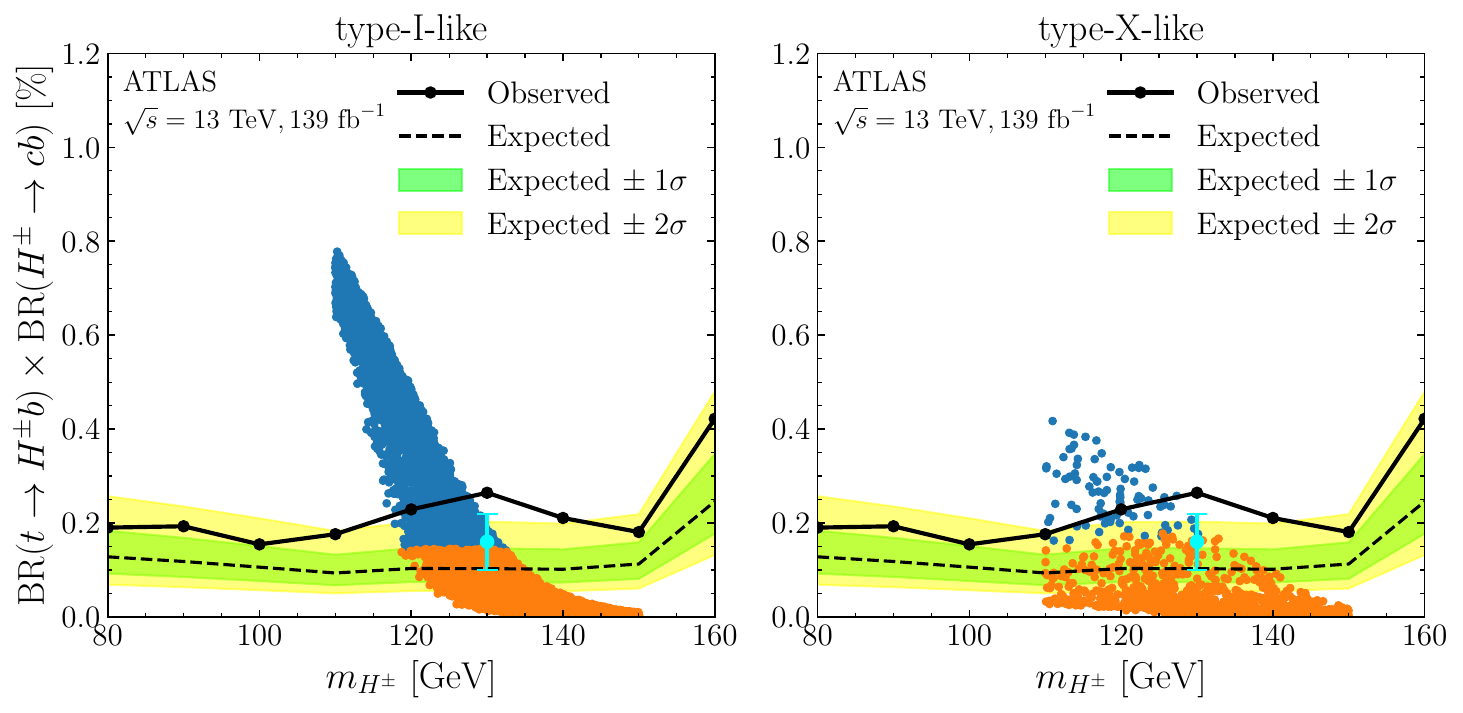}
				\includegraphics[scale=0.6]{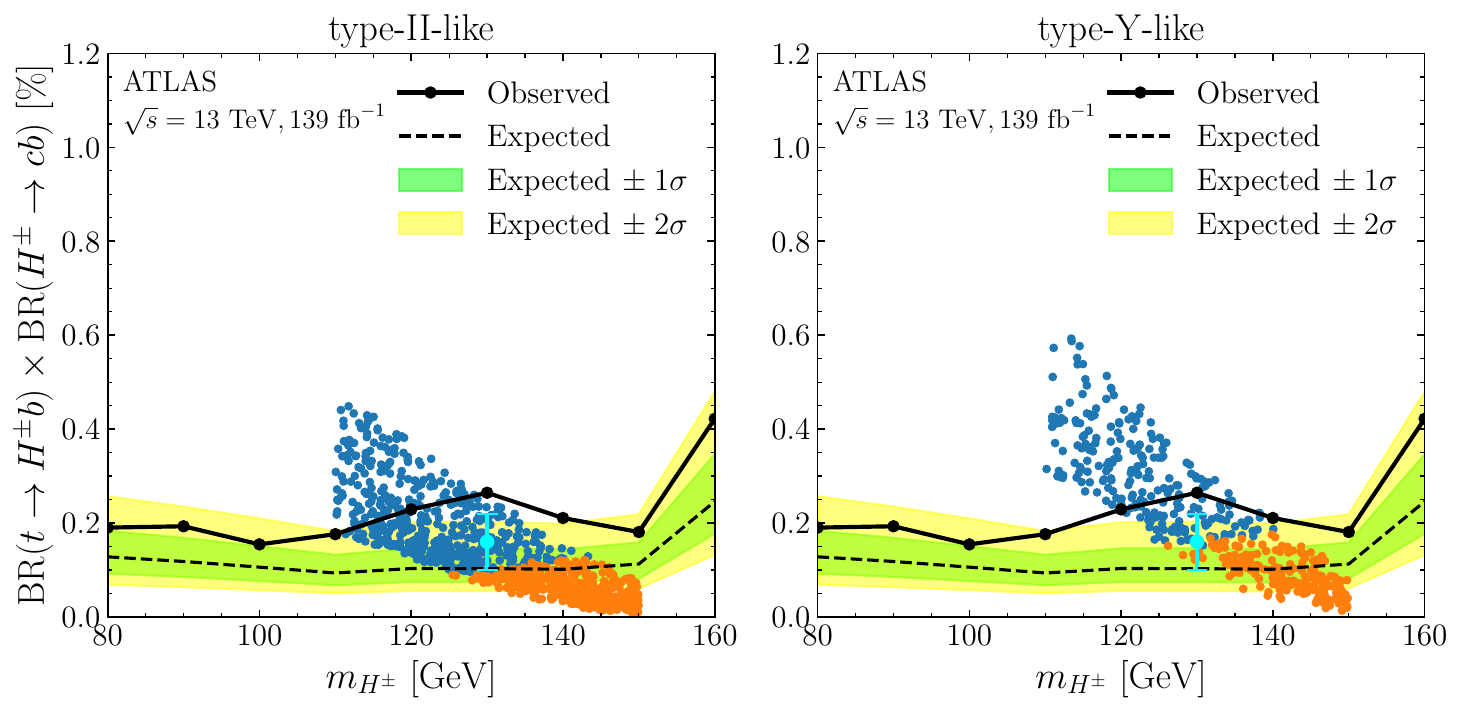}
				\caption{The product of branching ratios $\mathrm{BR}(t \rightarrow H^\pm b) \times \mathrm{BR}(H^\pm \rightarrow cb)$ as a function of charged Higgs mass for type-I-like (upper left panel), type-X-like (upper right panel), type-II-like (lower left panel) and type-Y-like (lower right panel). The observed and expected limits are shown by the black solid and dashed lines, respectively. The $1\sigma$ and $2\sigma$ uncertainties are represented by the light green and yellow bands, respectively. The cyan error bar shows the measured best-fit $\mathrm{BR}(t \rightarrow H^\pm b) \times \mathrm{BR}(H^\pm \rightarrow cb)$ and its uncertainty at 130 GeV. Orange (blue) points are allowed (excluded) by the constraint from $\mathrm{BR}(H^\pm \to cs+cb)$~\cite{CMS:2020osd}.} \label{fig1}
			\end{figure}	
			
			The parameter points passing the enforced constraints of type-I(X)-like and type-II(Y)-like are shown in the upper and lower panels, respectively, and indicated by orange colour. Recall that the main difference between type-I-like and type-X-like lies in the Higgs-lepton couplings. The same applies to type-II-like and type-Y-like. 
			The blue points are excluded by the CMS search for $H^\pm \rightarrow cs + cb$ \cite{CMS:2020osd} upon the further requirement that $\mathrm{BR}(t \rightarrow H^\pm b) \times \mathrm{BR}(H^\pm \to cs+cb)<0.3\%$. 
			\htb{It should be noted that Ref. \cite{CMS:2020osd} assumes that $\mathrm{BR}(H^\pm \to cs)=100\%$ and uses the $H^\pm \to cs$ labeling. Here, since branching ratio of  $\mathrm{BR}(H^\pm \to cb)$ is predicted to be comparable or even greater than the one of $H^+ \to cs$ (see Fig. \ref{BRHc_T1TX} and \ref{BRHc_T2TY}), we use the $H^\pm \to cs + cb$ labelling.}
			We also note that blue points didn't fail \texttt{HiggsBounds} test as the most sensitive search to these parameter points is the ATLAS search for $H^\pm$ in the $\tau\nu$ final state \cite{ATLAS:2018gfm}.\footnote{\texttt{HiggsBounds}~\cite{Bechtle:2013wla} uses the expected experimental limit to decide on the analysis exhibiting the highest statistical sensitivity to rule out the point in scrutiny, it then performs the exclusion test for the Higgs boson and analysis combination by computing the ratio of the model predictions to the observed experimental limit.} 
			One can read that both type-I-like and type-X-like can predict the best-fit value for the product of branching ratios $\mathrm{BR}(t \rightarrow H^\pm b)\times \mathrm{BR}(H^\pm \rightarrow cb)$ with $m_{H^\pm} = 130$ GeV (orange points), unlike in type-II-like where the limit on $H^\pm \rightarrow cs + cb$ excludes the parameter points that can explain such an excess. In the lower right panel, it can be seen that type-Y-like can accommodate the observed excess, predicting values at $1\sigma$ of the measured best fit $\mathrm{BR}(t \rightarrow H^\pm b)\times \mathrm{BR}(H^\pm \rightarrow cb)$ at $m_{H^\pm} = 130$ GeV. Note that in the type-Y realisation of the 2HDM with NFC a large branching ratio for $H^\pm \rightarrow cb$ can also be obtained for large $\tan\beta$ and $m_{H^\pm} < m_t$, accommodating the observed excess at 130 GeV \cite{ATLAS:2023bzb}. However, the $b \rightarrow s\gamma$ limit forbids this possibility. Such a scenario is indeed realised in the type-Y-like while agreeing with the $b \rightarrow s\gamma$ limit. \htb{Within this framework, $b\to s\gamma$ is mainly mediated by charged Higgs and  $W$ bosons, while extra contributions induced by off-diagonal terms are absent, therefore type-II(Y)-like contributions are similar to type-II(Y) with NFC plus extra terms relaxing $b \to s \gamma$ bound. We refer to Refs. ~\cite{Arhrib:2017yby,Hernandez-Sanchez:2012vxa} for further details.}
			
			We now turn to investigate the decay rates of a light charged Higgs in the mass range between 110 and 150 GeV. The expression of the partial width of a charged Higgs to light fermions, e.g. with mass below the top quark one, in the most general 2HDM are of the form, 				
			\begin{align}
			\Gamma(H^\pm\to u_i\bar d_j)&=\frac{3G_F m_{H^\pm}(m_{d_j}^2|X_{ij}|^2+m_{u_i}^2|Y_{ij}|^2)}{4\pi\sqrt 2},
			\label{width_ud} \\
			\Gamma(H^\pm\to {l}^\pm_j\nu_i)&=\frac{G_F m_{H^\pm} m^2_{{l}_j} |Z_{ij} |^2}{4\pi\sqrt 2},
			\label{width_tau}
			\end{align}
			where $X_{ij}$, $Y_{ij}$ and  $Z_{ij}$ are give by Eqs.~(\ref{Xij}), (\ref{Yij}) and (\ref{Zij}), respectively.
			
			\begin{figure}[t!]
				\centering
				\includegraphics[scale=0.6]{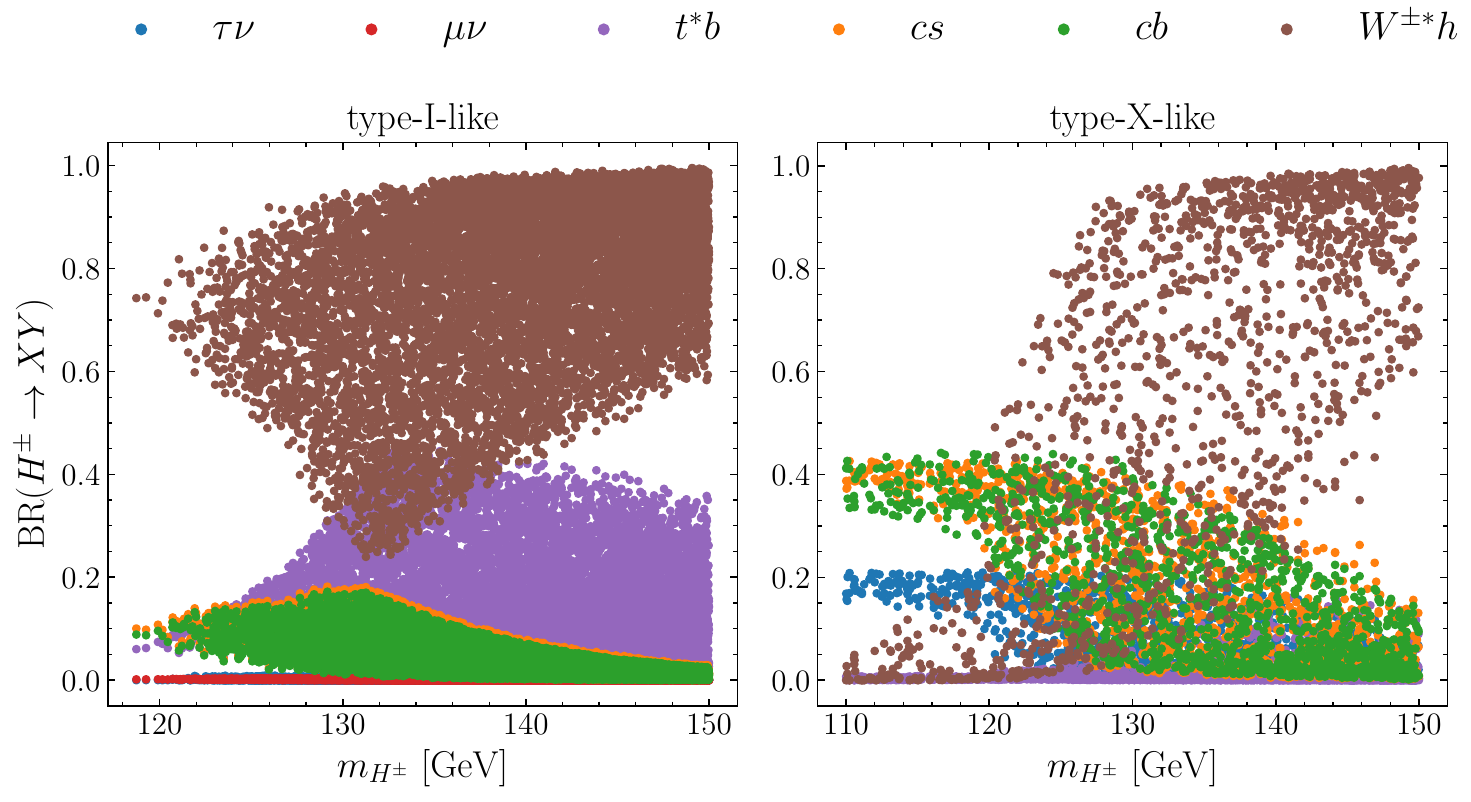}
				\caption{Branching ratios of the charged Higgs boson as a function of $m_{H^\pm}$ in 2HDM-III type-I-like (left panel) and type-X-like (right panel).} \label{BRHc_T1TX}
			\end{figure}
			
			Fig. \ref{BRHc_T1TX} displays the branching ratios of $H^\pm$ into different final states vs $m_{H^\pm}$ in type-I-like and type-X-like. \htb{Before reaching the mass region where the charged Higgs decay into $W^{\pm*}h$ starts to dominate over $cb$ and $cs$ decays (referred to as the $H^\pm \rightarrow W^{\pm*}h$ threshold)}, one can observe that the charged Higgs decay width in type-X-like is dominated by the $cs$ (orange points) and $cb$ (green points) final states with a branching ratio of 40\% followed by $\tau\nu$ ($\sim$20\%).  In type-I-like, the current upper limit on $cs+cb$ excludes the mass range below 120 GeV where the two channels, $cs$ and $cb$, compete and reach a significant branching ratio of 40\% (see Appendix~\ref{appen}, Fig~\ref{fig1_appendix}). This is because the $H^\pm cs$ and $H^\pm cb$ couplings enjoy an enhancement due to the new free parameters contribution ($\chi^d_{33}$, $\chi^u_{22}$), 
			\htb{
				\begin{eqnarray}
				\Gamma(H^\pm\to cb) \propto (m_{b}^2|X_{cb}|^2+m_{c}^2|Y_{cb}|^2), \quad  
				\Gamma(H^\pm\to cs) \propto (m_{s}^2|X_{cs}|^2+m_{c}^2|Y_{cs}|^2),  
				\end{eqnarray}
				$\text{with}  ~m_c Y_{cb} = V_{cb}m_c (Y - f(Y)/\sqrt{2} \chi_{22}^u ),~m_b X_{cb} =  V_{cb}m_b (X - f(Y)/\sqrt{2} \chi_{33}^d ),$  and
				$~m_c Y_{cs} = V_{cs}m_c (Y - f(Y)/\sqrt{2} \chi_{22}^u ),~m_s X_{cs} =  V_{cs}m_s (X - f(Y)/\sqrt{2} \chi_{22}^d )$.} If these two parameters vanish, one can then restore the 2HDM (type-I and type-X) with NFC in which such couplings would be suppressed by the parameter $1/\tan\beta$ and therefore charged Higgs decays in that mass region would be dominated by $\tau\nu$ final state. 
			
			Above the  $H^\pm \rightarrow W^{\pm*}h$ threshold, the charged Higgs decay into $W^{\pm*}h$ (brown points) is dominant over the parameter space in both type-I and type-X. The large $\mathrm{BR}(H^\pm \rightarrow W^{\pm*}h)$ is attributed to the $H^\pm W^\mp h$ vertex, which is given by the parameter $\cos(\beta - \alpha)~(\approx 1)$. The branching ratios BR($H^\pm \rightarrow W^{\pm*}H$) and BR($H^\pm \rightarrow W^{\pm*}A$) are suppressed by $\sin(\beta - \alpha)$ and kinematics, respectively.
			As shown in the figure, the decay into $t^* b$ (purple points) could reach $\sim$ 40\% in type-I-like, while other decay channels are suppressed, whereas in type-X-like, additional channels, including $cs$, $cb$, and $\tau\nu$, contribute to the overall decay width.
			
			\begin{figure}[t!]
				\centering
				\includegraphics[scale=0.6]{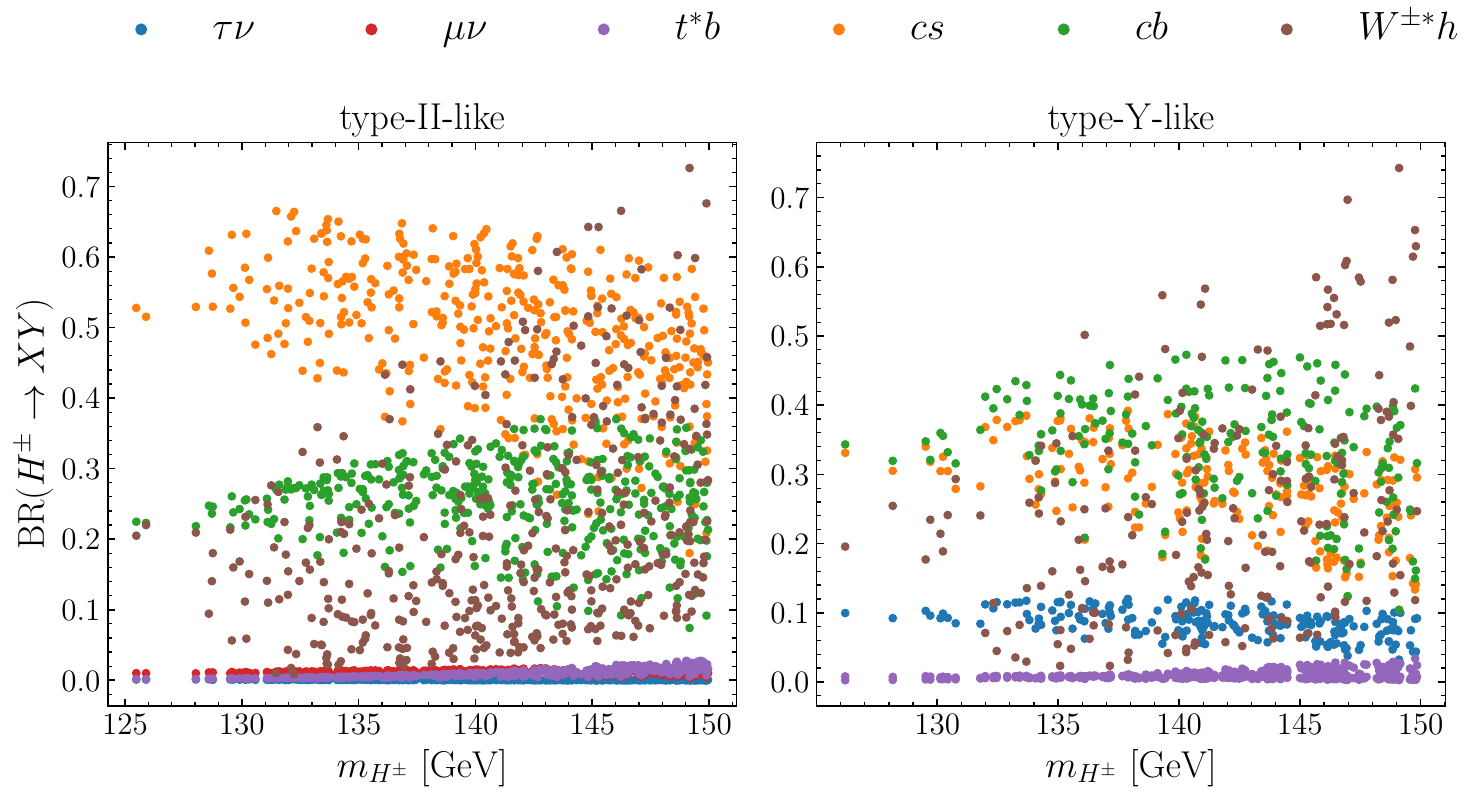}
				\caption{Branching ratios of the charged Higgs boson as a function of $m_{H^\pm}$ in type-II-like (left panel) and type-Y-like (right panel).} \label{BRHc_T2TY}
			\end{figure}   
			The results of the 2HDM type-II-like and type-Y-like are shown in Fig~\ref{BRHc_T2TY}. We see that the decay into $cs$ (orange points) dominates in all mass ranges of $H^\pm$ in type-II-like. The decay into $cb$ (green points) is also significant with the branching ratio reaching values up to 40\%. These final states enjoy enhancement from $\tan\beta$ and $\chi^d_{33}$ as well as $\chi^u_{22}$. 
			%This underscores the fact behind predicting the measured best-fit BR$(t \rightarrow H^+ b)\times\mathrm{BR}(H^+ \rightarrow c\bar b)$ at $\pm 2\sigma$ (see Fig. \ref{fig1}).
			The $W^{\pm*}h$ decay is slightly competing around 140 GeV or so and one should expect it to be dominant above $m_{H^\pm} > 150$ GeV.
			\htb{$H^\pm \rightarrow \tau\nu$ and $H^\pm \to \mu\nu$ are negligible with branching ratios reaching at most 0.16\% and 1.74\%, respectively.
				In some cases, $H^\pm$ could be leptophobic when its coupling to leptons approaches zero, resulting in branching ratios $\ll 1\%$ \cite{Hernandez-Sanchez:2012vxa}.}
			The $t^* b$ decay is also negligible.  
			While in type-Y-like (see right panel of Fig. \ref{BRHc_T2TY}), for charged Higgs masses below 135 GeV (the $H^\pm \rightarrow W^{\pm*}h$ threshold), the $H^\pm$ decays are dominated by the $cb$ channel, with branching ratio reaches values above 40\%, competing with $cs$ channel. The $\tau\nu$ decay could reach values up to 10\%. 
			Other decays especially $\mu\nu$ and $t^*b$ are found to be negligible. 
			\htb{It is worth noting that the observed difference in density of points is mainly due to experimental constraints, especially BSM Higgs searches at collider experiments and measured properties of the SM-like Higgs.}

			Let us now explore the emerging signatures that could be useful to identify a light charged Higgs boson at the LHC. As seen above, the $H^\pm \rightarrow cs$, $cb$ and $W^{\pm*}h$ decays are important, depending on a specific Yukawa texture. To produce the charged Higgs boson, we use the typical $H^\pm$ production mode, i.e. the top quark pair production and decay $pp \rightarrow t\bar t, t \rightarrow H^\pm b$. We then multiply the production cross section, computed at $\sqrt{s} = 14$ TeV,\footnote{We use the predicted $t\bar t$ production cross section, $\sigma(pp \rightarrow t\bar t) = 985.7$ pb, calculated at next-to-next-to-leading order in QCD, from the webpage \url{https://twiki.cern.ch/twiki/bin/view/LHCPhysics/TtbarNNLO}.} by the appropriate branching ratios. The results are depicted in Figs. \ref{tbHc_T1TX} and \ref{tbHc_T2TY}.
			
			\begin{figure}[t!]
				\centering
				\includegraphics[scale=0.6]{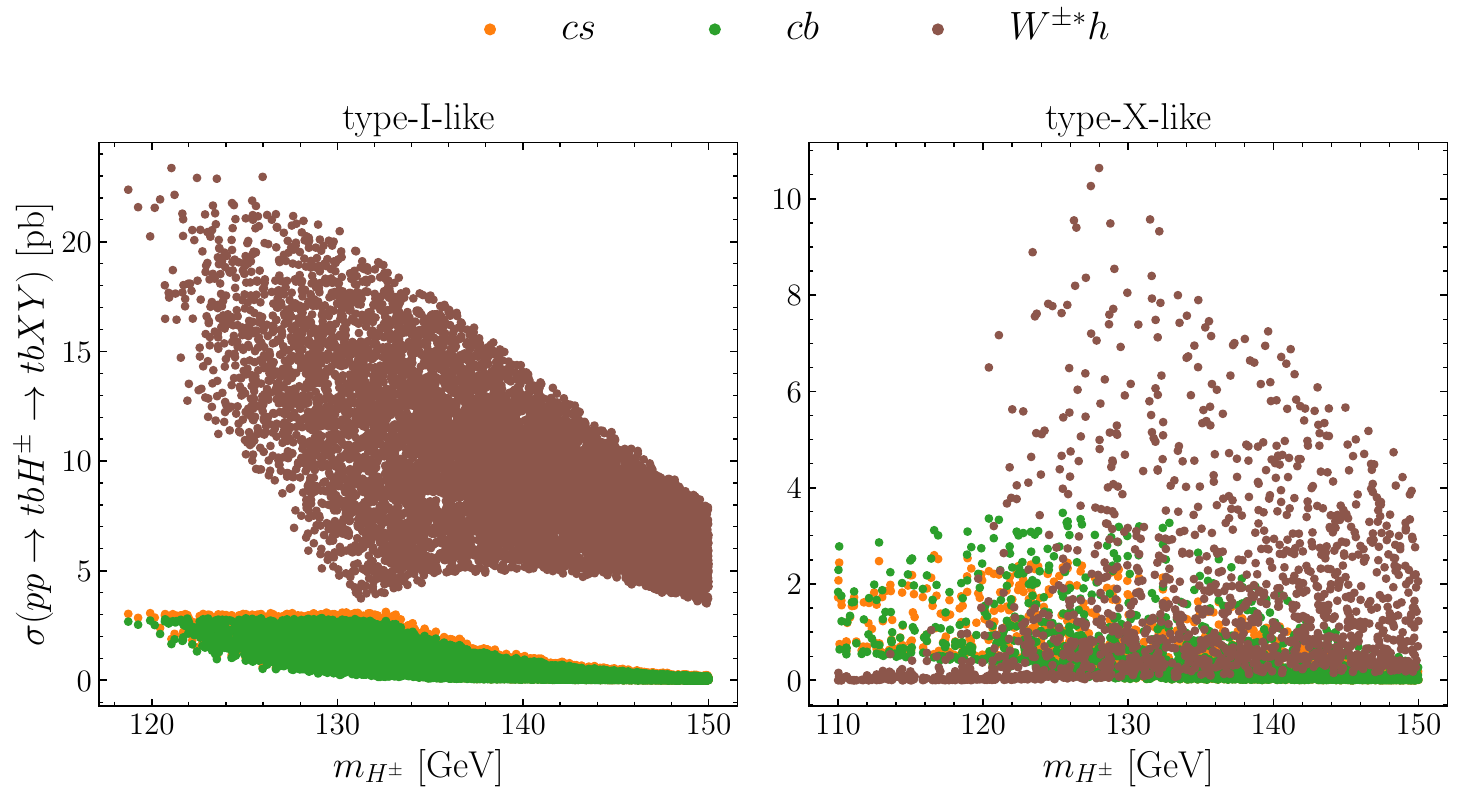}
				\caption{Signal cross sections for $pp \rightarrow tbH^\pm \rightarrow tbcs$ (orange points) $tbcb$ (green points) and $tbW^{^\pm*}h$ (brown points) as a function of the charged Higgs boson mass, $m_{H^\pm}$. The type-I-like (type-X-like) results are shown in the left (right) panel.} \label{tbHc_T1TX}
			\end{figure}
			
			We show in Fig. \ref{tbHc_T1TX} the total cross section for the $pp \rightarrow tbH^\pm \rightarrow tbcs$ (orange points) $tbcb$ (green points) and $tbW^{\pm*}h$ (brown points) processes as a function of $m_{H^\pm}$. The type-I-like predicted results are depicted in the left panel while the type-X ones are shown in the right panel. For charged Higgs masses close to 130 GeV, the $tbW^{\pm*}h$ final state yields a significant cross section reaching values above 20 pb. Considering the branching ratio of $h \rightarrow b\bar b$,  which could reach values up to 99\%, searches for light $H^\pm$ in this final state could be interesting.\footnote{In type-I of the 2HDM with NFC, the charged Higgs boson decay $H^\pm \rightarrow W^{\pm(*)} h$ and/or $H^\pm \rightarrow W^{\pm(*)} A$ could be a promising discovery mode for $H^\pm$ \cite{Arhrib:2020tqk,Bahl:2021str,Arhrib:2021xmc,Wang:2021pxc,Arhrib:2021yqf,Arhrib:2022inj,Krab:2022lih,Sanyal:2023pfs,Li:2023btx}.} The $tbcs$ and $tbcb$ final states dominate for lower charged Higgs masses below 120 GeV. \htb{For charged Higgs boson masses above $\sim 125$ GeV, the $tbt^*b$ final state dominates over $tbcs$ and $tbcb$ final states and reaches values up to $\sim 9$ pb around $m_{H^\pm} \sim 133$ GeV.} The type-X-like shows nearly the same behaviour. For lower $m_{H^\pm}$, the $tbcb$ and $tbcs$ signatures dominate with production cross sections reaching a level of $3$ pb. Close to $m_{H^\pm} = 130$ GeV, the emerging $tbW^{\pm*}h$ signature becomes important with cross section reaching values up to 10 pb. 
			In type-I-like and/or type-X-like, the $H^\pm \rightarrow W^{\pm*}h(\rightarrow b\bar b)$ decay channel could also be useful to confirm or refute the ATLAS first hint of a light charged Higgs boson with a mass close 130 GeV.
			
			The discussed final states ($tbcs$, $tbcb$ and $tbW^{\pm*}h$) have different behaviour in type-II-like and type-Y-like. This is demonstrated in the left (type-II-like) and right (type-Y-like) panels of Fig. \ref{tbHc_T2TY}. As expected, the $tbcs$ final state is especially dominant. Close to 130 GeV, its cross section lies up to 4 pb while the cross section for $tbcb$ could reach values above 2 pb. Here,  unlike type-I-like (and type-X-like), the $tbW^{\pm*}h$ signature is comparable to $tbcb$ one. A similar observation is seen in type-Y-like, except that $tbcb$ is dominating (followed by $tbcs$), especially for charged Higgs boson masses below 130 GeV. Above 130 GeV, the $tbW^{\pm *}h$ final state becomes important and can dominate for large mass values. 
			%Here, based on these Yukawa textures, $H^\pm \rightarrow cs$ might also be useful to further search for light $H^\pm$. 
			
			\begin{figure}[t!]
				\centering
				\includegraphics[scale=0.6]{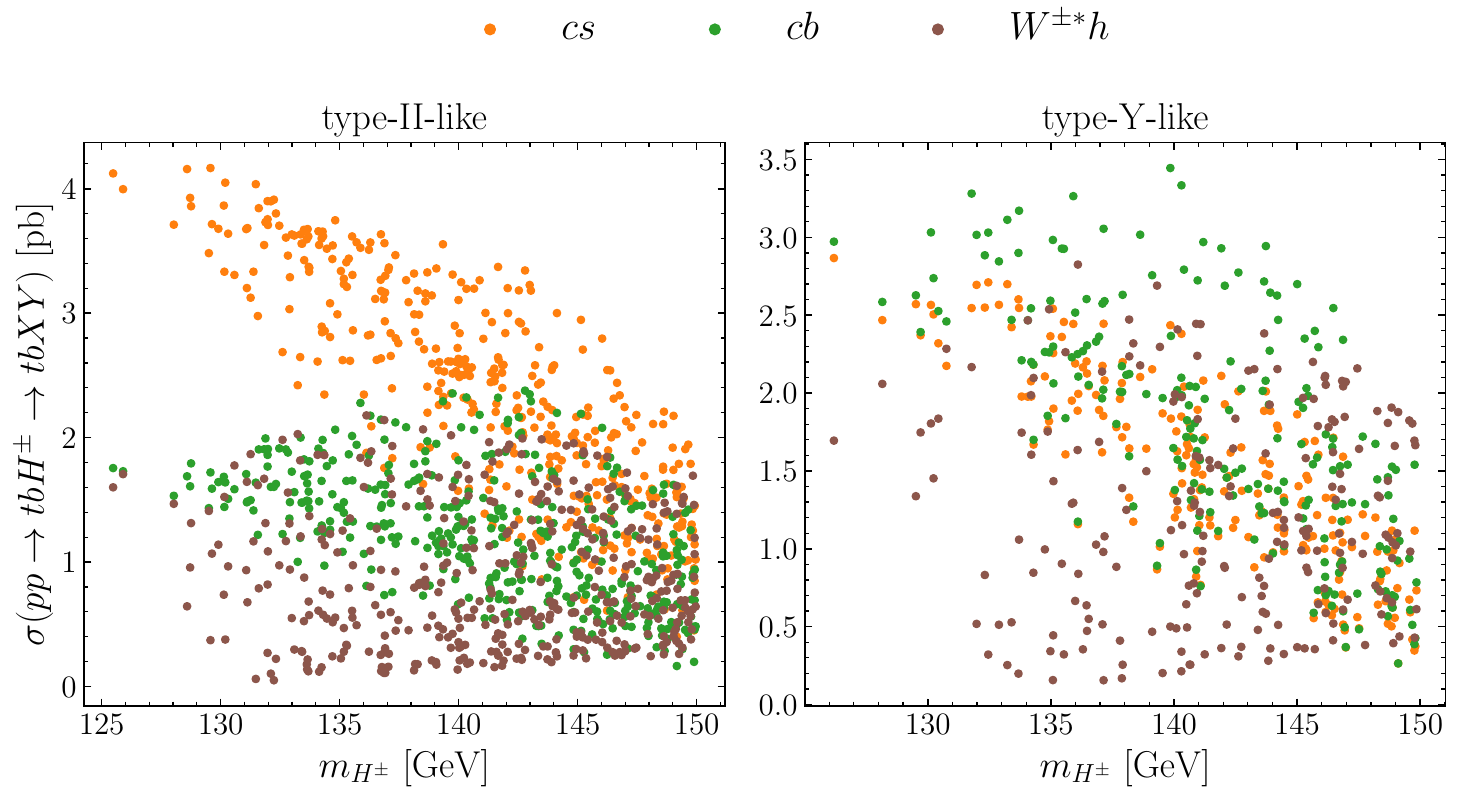}
				\caption{Signal cross sections for $pp \rightarrow tbH^\pm \rightarrow tbcs$ (orange points) $tbcb$ (green points) and $tbW^{^\pm*}h$ (brown points) as a function of the charged Higgs boson mass, $m_{H^\pm}$. The type-II-like (type-Y-like) results are shown in the left (right) panel.} \label{tbHc_T2TY}
			\end{figure}
			
			Finally, we point out that using top pair ($t\bar{t}$) production at the LHC, where one top quark decays into a charged Higgs boson and a bottom quark and the other top decays into a $W$-boson and a $b$-quark, we anticipate much larger event rates, specifically for the final states $csW bb$, $cbWbb$ and $W^{*}hWbb$. In the latter, $h$ would dominantly decay into a pair of $b$ quarks. \htb{Note that even in type-X-like, $h\to b\bar{b}$ dominates over $h\to \tau^+\tau^-$ in the parameter space due to the small $\tan\beta$. The latter is strongly limited by experimental constraints, particularly flavour physics constraints.} leading to $W^*Wbbbb$ final state, where $W^*$ being off-shell decaying into a lepton\footnote{The leptonic decay of the $W$-boson could provide a useful lepton trigger. The other $W$ (on-shell) would decay hadronically into a pair of quarks.} (electron or muon) and a neutrino, thus giving rise to a significant number of events, e.g. $N\sim L\times \sigma \sim 900000(450000)$ in type-I(X)-like near $m_{H^\pm} = 130$ GeV, where $L = 300$ fb$^{-1}$ is the integrated luminosity. For this signature, the emerging lepton might be soft. Therefore, the charged Higgs boson can be reconstructed using the soft lepton, the missing transverse energy (comes from neutrino) and the reconstructed $h$. The mass of the latter can be reconstructed by finding the appropriate combination of bottom quarks that peaks near $m_h$. The $\ell\nu bbbbqq$ final state could be promising to further search for evidence of $H^\pm$ at the LHC. Investigating the possibility of observing such a prominent signature would be considered for future study. 
			%The $\ell\nu bbbbqq$ final state could be promising to further search for evidence of $H^\pm$ at the LHC.
			%The study of the LHC discovery potential of the $\ell\nu bbbbqq$ final state is postponed for future work.
			%The associated backgrounds (especially $t\bar t$) would be suppressed using the same method as in Ref. \cite{Li:2023btx}.}    

			\section{Conclusion}
			ATLAS experiment has observed for the first time, using the data set collected at a center-of-mass energy of 13 TeV, a local excess of $3\sigma$ at $130$ GeV during the search for a light charged Higgs boson decaying into a charm quark and a bottom quark \htb{\cite{ATLAS:2023bzb}}. This excess is best fitted by a product of branching ratios BR$(t \rightarrow H^+ b)\times\mathrm{BR}(H^+ \rightarrow c\bar b) = 0.16\% + 0.06\%$. Such excess is of particular interest since it can be useful to investigate any BSM predicting a charged Higgs state in its scalar sector. 
			
			Focusing on the 2HDM type-III, we have demonstrated the existence of a viable parameter space that could accommodate the observed excess at a charged Higgs mass of 130 GeV, while being in good agreement with various theoretical constraints and current experimental limits.  We have then studied the $H^\pm$ phenomenology in the four Yukawa textures of the 2HDM type-III in the excess region with $m_{H^\pm} \in [110,~150]$ GeV.      
			We have demonstrated that the decay channels $H^\pm \rightarrow cs$ and $H^\pm \rightarrow W^{\pm*}h$ could be interesting and might be used to confirm or refute the first hint of light $H^\pm$ reported by the ATLAS experiment, where the latter channel relies on a discovery of a new light neutral Higgs-like scalar. Finally, we have studied three charged Higgs signals at the LHC that could further probe light $H^\pm$. 
			
			\section*{Acknowledgements}
			AA is supported by the Moroccan Ministry of Higher Education and Scientific Research MESRSFC and CNRST: Projet PPR/2015/6.
			MK thanks the support of grant NSTC 111-2639-M-002-004-ASP of Taiwan
			and acknowledges the use of CNRST/HPC-MARWAN in completing this work. SS is fully supported through the NExT Institute.

			\bibliographystyle{JHEP}
			\bibliography{references}
			
			\newpage
			\appendix
			\section{Appendix}
			\label{appen}

			\begin{figure}[h!]
				\centering
				\includegraphics[scale=0.6]{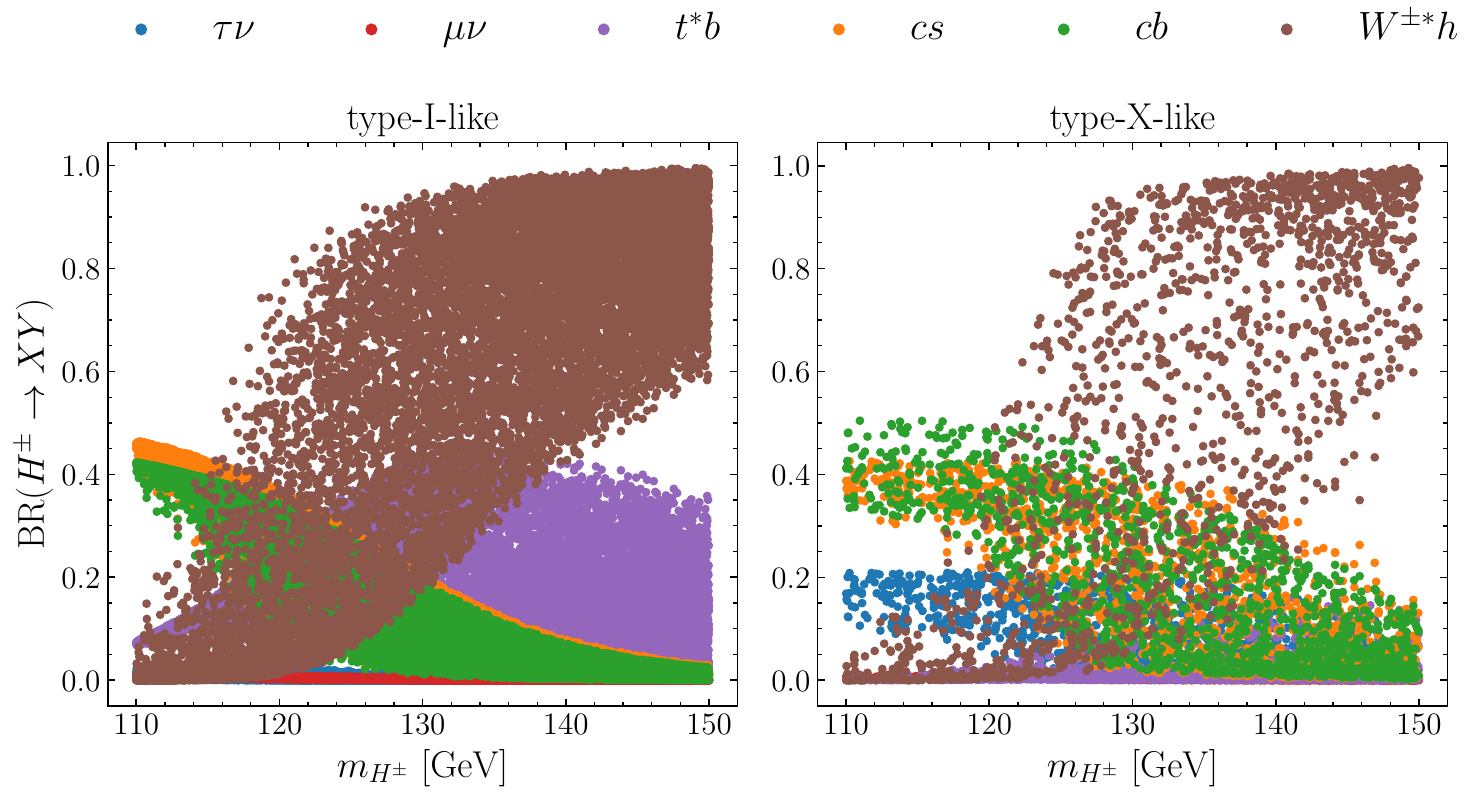}
				\includegraphics[scale=0.6]{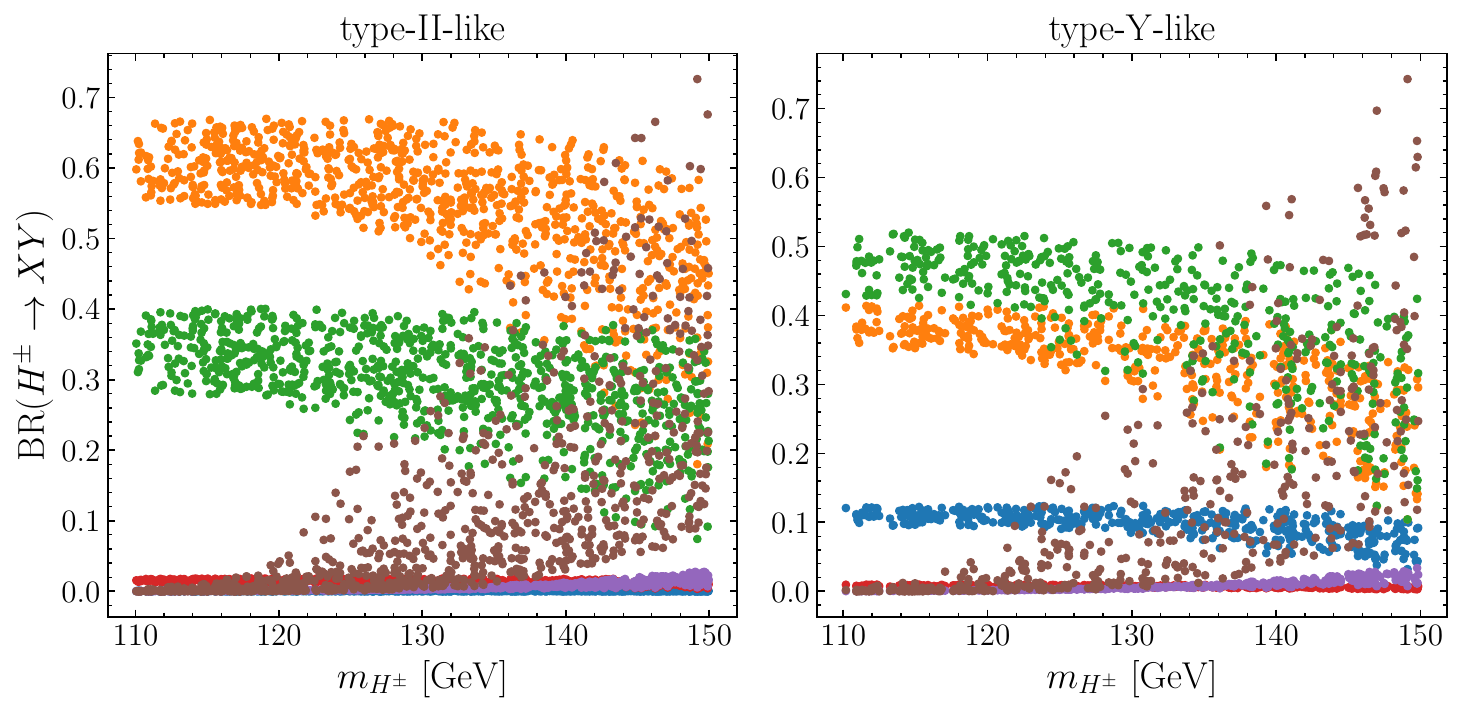}
				\caption{Branching ratios of the charged Higgs boson as a function of $m_{H^\pm}$ in 2HDM-III type-I-like (upper left panel), type-X-like (upper right panel), type-II-like (lower left panel) and type-Y-like (lower right panel). Only the exclusion limits from \texttt{HiggsBounds} are applied here. The CMS limit from $H^\pm \rightarrow cs+cb$ \cite{CMS:2020osd} is not further enforced to show the potential impact of this limit on the allowed parameter space in 2HDM type-III.}
				\label{fig1_appendix}
			\end{figure}
		\end{document}